\documentclass{iaus}
\usepackage{graphicx}

\title[Recurrent Novae: an X-ray Perspective] 
{The White Dwarf Mass and the Accretion Rate of Recurrent Novae: \\
 an X-ray Perspective}
\author[K. Mukai et al.]   
{Koji Mukai$^{1,2}$, Jennifer L. Sokoloski$^3$, Thomas Nelson$^4$, \and Gerardo J. M. Luna$^5$}

\affiliation{$^1$CRESST and X-ray Astrophysics Laboratory, NASA Goddard Space Flight Center, Greenbelt, MD 20771, USA \\ email: {\tt Koji.Mukai@nasa.gov} \\[\affilskip]
$^2$Also Department of Physics, University of Maryland Baltimore County,
    1000 Hilltop Circle, Baltimore, MD 21250, USA\\
$^3$Columbia Astrophysics Laboratory, 550 W. 120th St., 1027 Pupin Hall,
    Columbia University, New York, NY 10027, USA \\
$^4$School of Physics and Astronomy, University of Minnesota,
    115 Church St. SE, Minneapolis, MN 55455, USA \\
$^5$Instituto de Ciencias Astron\'omicas, de la Tierra y del Espacio
    (ICATE/FCEFyN), Av. Espa\~na Sur 1512, J5402DSP, San Juan, Argentina }

\pubyear{2011}
\volume{281}  
\pagerange{119--126}
\jname{Title of your IAU Symposium}
\editors{A.C. Editor, B.D. Editor \& C.E. Editor, eds.}
\begin{document}

\maketitle

\begin{abstract}
We present recent results of quiescent X-ray observations of recurrent
novae (RNe) and related objects.  Several RNe are luminous hard X-ray
sources in quiescence, consistent with accretion onto a near Chandrasekhar
mass white dwarf.  Detection of similar hard X-ray emissions in old novae
and other cataclysmic variables may lead to identification of additional
RN candidates.  On the other hand, other RNe are found to be comparatively
hard X-ray faint.  We present several scenarios that may explain this
dichotomy, which should be explored further.
\end{abstract}

\firstsection 
\section{Introduction}

By definition, a recurrent nova (RN) has been seen to undergo multiple
episodes of thermonuclear runaway within the last century or so.  For
the hydrogen-rich envelope to reach the high temperature and density
required for a runaway in such a short period, the white dwarf must
be massive and its accretion rate must be high.  Although only ten
Galactic RNe are currently known, the true number of
RNe is likely to be much larger, considering the low discovery probability
of nova outbursts \cite[(Schaefer 2009)]{S2010}.  Two important
goals for RN observers in the context of Type Ia progenitors therefore are
(i) observational determination of white dwarf mass and accretion rate;
and (ii) search for hitherto undiscovered or unrecognized RNe.

In non magnetic cataclysmic variables (CVs) and symbiotic stars,
X-rays are emitted in the boundary layer between the disk and the
white dwarf.  Optically thin boundary layers predominantly emit
hard X-rays; even optically thick boundary layers are seen to retain
some hard X-ray flux, presumably because the surface layer
remains optically thin \cite[(Patterson \& Raymond 1985)]{PR1985}.
These hard X-rays are multi-temperature plasma emission whose maximum
temperature is strongly constrained by the depth of the gravitational
potential, i.e., the white dwarf mass.  Thus, hard X-ray observations
may be a viable alternative method to optical and UV spectroscopy in
our study of white dwarf masses, particularly in cases of high
interstellar extinction.

If the Keplerian flow just above the white dwarf surface is strongly
shocked, then the shock temperature is half of the free-fall case,
well known in the studies of magnetic CVs.  Multiple groups have
used X-ray spectroscopy to infer the white dwarf mass in magnetic CVs;
the study of quiescent X-rays from dwarf novae
\cite[(Byckling \etal\ 2010)]{Bea2010} suggests that this is also possible
for non-magnetic CVs.  One complication is that the hard X-ray
emission of a dwarf nova usually becomes fainter and softer during
outburst when a part of the boundary layer becomes optically thick
(see below for a possible reason).  Even so, the maximum
temperature derived for the hard component sets a firm lower limit
for the white dwarf mass.

\section{X-ray Bright RNe}

\begin{figure}[t]
\begin{center}
 \includegraphics[width=10.5 cm]{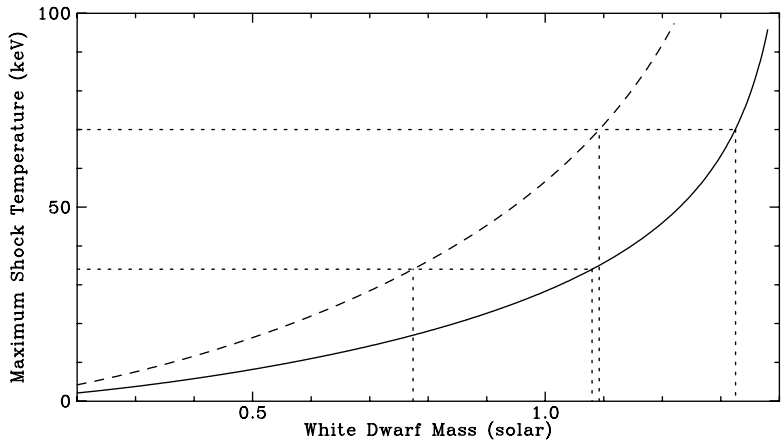} 
 \caption{The expected maximum temperature of the optically thin X-rays
for the magnetic (dashed line) and non-magnetic (solid) cases.  The lower
limit of 34 keV for V2487 Oph corresponds to 0.77 M$_\odot$ (magnetic)
or 1.08 M$_\odot$ (non-magnetic) white dwarf.  The best-fit temperature
of 70 keV corresponds to 1.01 M$_\odot$ and 1.33 M$_\odot$, respectively.}
   \label{fig1}
\end{center}
\end{figure}

In recent years, four symbiotic stars have been detected as luminous hard
X-ray sources in the {\sl Swift\/} BAT and {\sl INTEGRAL} surveys
\cite[Kennea \etal\ (2009)]{Kea2009}.  One of the four is T~CrB,
with a 15--150 keV luminosity of 7$\times 10^{33} \times [d/1kpc]^2
$ erg\,s$^{-1}$.  The current generation of hard X-ray all-sky surveys 
have a detection limit of order $\sim 10^{-11}$ ergs\,cm$^{-2}$s$^{-1}$,
or $\sim 10^{33}$ erg\,s$^{-1}$ at 1 kpc, below which many more hard X-ray
bright symbiotics are likely to exist.  In fact,
\cite[Luna \etal\ (2010)]{Lea2010} have
significantly increased the number of known hard X-ray sources
among symbiotic stars through pointed observations with {\sl Swift\/}
XRT.  

\cite[Kennea \etal\ (2009)]{Kea2009} made the case that T~CrB contains
a near Chandrasekhar mass, non-magnetic white dwarf using a Bremsstrahlung
fit to the BAT spectrum available at the time.  We have updated their
argument as follows.  We fit the BAT spectrum below 100 keV from the
{\sl Swift\/} BAT 58-month survey with a cooling flow model, and
obtain the maximum temperature of kT$_{max} = 46\pm$6 keV.  If the
emission is from an optically thin boundary layer, it implies
a white dwarf mass of M$_{wd}$=1.2 M$_\odot$.  More likely, a significant
portion of the boundary layer in T~CrB is optically thick, given that the
high UV luminosity \cite[(Selvelli \etal\ 1992)]{Sea1992}.  If this
has resulted in the reduction of kT$_{max}$ by a factor of 1.7,
as it does in SS~Cyg in outburst, then  M$_{wd}$=1.35 M$_\odot$
in T~CrB.

V2487 Oph is the first nova for with a pre-outburst X-ray detection
\cite[(Hernanz \& Sala 2002)]{HS2002}.  It is also a hard X-ray source
detected in the {\sl INTEGRAL\/} and {\sl Swift\/} BAT surveys.
Although this is a CV and not a symbiotic system (the mass donor in
V2487~Oph is not a red giant), its X-ray spectrum can be compared to
that of T~CrB.  Using the BAT 58-month survey data and
a cooling flow model, the best fit kT$_{max}$ is 70 keV; the 90\%
lower limit is 34 keV, while the upper limit cannot be constrained
due primarily to the limited grid of plasma models available within
the cooling flow model.  For a distance of 12 kpc
\cite[(Schaefer 2010)]{S2010}, its 2--10 keV luminosity is close to
10$^{35}$ ergs\,s$^{-1}$.  In contrast, intermediate polars (IPs),
the most hard X-ray luminous subclass of magnetic CVs, usually do not
exceed 10$^{34}$ ergs\,s$^{-1}$.  V2487~Oph was established as an RN
by \cite[Pagnotta \etal\ (2009)]{Pea2009}, who discovered its 1900
outburst in photographic plates.

V2487 Oph is often considered a candidate IP, based on its X-ray
appearances.  However, the signature of a spin period, a key defining
characteristic of IPs, has never been seen in this object.  Moreover,
the hardness of its X-ray spectrum can be explained by two scenarios:
it may be a magnetic CV with a moderately massive white dwarf, or
it may be a non-magnetic CV with a near Chandrasekhar mass white dwarf
(Figure\,1).  Also considering the fact that V2487~Oph is a RN,
the latter interpretation is very attractive.

V2491~Cyg has several similarity with V2487~Oph, including a pre-nova
X-ray detection \cite[(Ibarra \etal\ 2009)]{Iea2009}
and the strong post-nova X-ray emission \cite[(Takei \etal\ 2011)]{Tea2011}.
These X-ray characteristics alone make it a candidate RN, which seems
to be corroborated by several RN-like characteristics in the optical.

We also find several other CVs that are not known to be a nova, let
alone RNe, and have not been firmly established to be magnetic,
in the BAT 58-month survey catalog: AH~Men, V426~Oph, TW~Pic, and
V1082~Sgr (two dwarf novae, SS~Cyg and RU~Peg, are also detected by BAT
but their accretion rates are presumably too low to be of interest
in this context).  These 4 systems may deserve further attention.

\section{X-ray Faint RNe}

However, it is now clear that not all RNe are X-ray bright.  RS~Oph
is significantly fainter in X-rays (2$\times 10^{33}$ ergs\,s$^{-1}$,
0.3--10 keV on day 538) and significantly softer (kT$_{max} \sim 5$ keV
in a cooling flow fit) than T~CrB \cite[(Nelson \etal\ 2011)]{Nea2011}.
T~Pyx in quiescence is also a faint (and not supersoft) X-ray source.
The central binary has a luminosity of $\sim 10^{32}$ ergs\,s$^{-1}$
for a distance of d=3.5 kpc \cite[(cf. Balman 2010)]{B2010}.

We have observed 3 additional Galactic RNe, V394~CrA, CI~Aql, and
and IM~Nor with {\sl XMM-Newton\/}.  For IM~Nor, there is a source
at RA=15:39:27.6, Dec=$-$52:18:55.1, roughly 30$"$ from the cataloged
position of this RN, at about 4.2$\times 10^{-3}$ c\,s$^{-1}$ with
a poorly constrained spectrum.  If this X-ray source is the RN,
its luminosity is roughly 5 $\times 10^{30} [d/2kpc]^2$ ergs\,s$^{-1}$
(2--10 keV).  V394~CrA and CI~Aql are undetected with 2--10 keV
upper limits of 5 $\times 10^{31} [d/5kpc]^2$ ergs\,s$^{-1}$
and 5 $\times 10^{30} [d/2kpc]^2$ ergs\,s$^{-1}$, respectively.
We note that IM~Nor and CI~Aql are known to be eclipsing.  If
the inclination angles are high enough, or the disks thick enough,
it is possible that the white dwarf is always hidden from our
view, in which case the observations do not necessarily reflect
the true X-ray luminosity of these systems.

Nevertheless, it is clear that not all RNe are luminous hard
X-ray sources like T~CrB and V2487~Oph.  These X-ray faint
RNe may nevertheless harbor massive white dwarfs accreting
at high rates.  The X-ray luminosity can be reduced if the
boundary layer is completely optically thick, if the white
dwarf is rotating at near break-up spin, or if the optically
thin part of the boundary layer is Compton-cooled.

When Bremsstrahlung is the primary cooling mechanism of
the post-shock plasma, the cooling time is inversely proportional
to the square of the density; the plasma cools and becomes denser,
which increases the cooling efficiency, and the cycle repeats.
However, if there is a strong external field of soft photons, Compton
cooling may be more efficient than Bremsstrahlung in parts of the boundary
layer.  \cite[Nelson \etal\ (2011)]{Nea2011} explored the possibility of Compton cooling for
RS~Oph, and \cite[Fertig \etal\ (2011)]{Fea2011} applied the same idea
to dwarf novae in outburst.  Compton cooling dominates over Bremsstrahlung
only in the less dense, which is also the hotter, regions of the boundary
layer.  Therefore, Compton cooling lowers both the temperature and the
luminosity of the observed hard X-ray emission.  This is indeed seen in
dwarf novae in outburst, in which the seed photons are supplied by the
optically thick part of the boundary layer.  If, in RS~Oph, the entire
white dwarf is still hot and luminous ($\sim 10^{35}$ ergs\,s$^{-1}$),
this can provide a copious amount of seed photons.  Therefore, the
potential difference in the white dwarf temperature can  explain
the different X-ray properties of RS~Oph and T~CrB in principle.

The luminosity of the white dwarf in RNe is likely to depend
on the time since last outburst, somewhat longer-term history
of recent outbursts, and other factors.  If our Compton cooling
model is correct, then RNe with luminous white dwarf are X-ray
faint.  While this interpretation is not firmly established,
it offers a plausible explanation for the diverse
X-ray characteristics of quiescent RNe without invoking
a low mass white dwarf or a low accretion rate.

\section{Conclusions}

We have learned that some quiescent RNe are luminous, hard X-ray sources.
Although the fraction of such systems among RNe is unknown, this offers
a new method for discovering RNe candidates.  A sensitive hard X-ray
all-sky survey, such as expected using
the {\sl e-ROSITA\/} mission, will be very useful in this regard.

The X-ray spectra of T~CrB and V2487~Oph are consistent with what
one would expect for an optically thin, or partially optically thick,
boundary layer around a massive non-magnetic white dwarf.  With the
current level of knowledge, we cannot measure the white dwarf mass
accurately, although high temperature X-ray emission requires high
M$_{wd}$.  The X-ray luminosity is a direct measure of the accretion
rate onto the white dwarf if and only if the boundary layer is
completely optically thin; in the partially optically thick case,
we can only provide a lower limit for the accretion rate.

The existence of X-ray faint RNe requires an explanation.  We offer
Compton cooling as a possibility.  In addition to more quantitative
exploration of this process, we must understand the evolution of
the white dwarf luminosity appropriate in the RN regime.


\begin{thebibliography}{}

\bibitem[Balman (2010)]{B2010}
{Balman, S.} 2010, \textit{MNRAS}, 404, L26

\bibitem[Byckling \etal\ (2010)]{Bea2010}
{Byckling, K., Mukai, K., Thorstensen, J.R., \& Osborne, J.P.} 2010,
\textit{MNRAS}, 408, 2298

\bibitem[Fertig \etal\ (2011)]{Fea2011}
{Fertig, D., Mukai, K., Nelson, T., \& Cannizzo, J.} 2011,
\textit{PASP}, in press

\bibitem[Hernanz \& Sala (2002)]{HS2002}
{Hernanz, M., \& Sala, G.} 2002, \textit{Science}, 298, 383

\bibitem[Ibrra \etal\ (2009)]{Iea2009}
{Ibarra, A., Kuulkers, E., Osborne, J.P., Page, K., Ness, J.U., Saxton, R.D.,
Baumgartner, W., Beckmann, V., Bode, M.F., Hernanz, M., Mukai, K., Orio, M.,
Sala, G., Starrfield, S., \& Wynn, G.A.} 2009, \textit{A\&A}, 497, L5


\bibitem[Kennea \etal\ (2009)]{Kea2009}
{Kennea, J.A., Mukai, K., Sokoloski, J.L., Luna, G.J.M.,
Tueller, J., Markwardt, C.B., \& Burrows, D.N.} 2009, \textit{ApJ}, 701, 1992

\bibitem[Luna \etal\ (2010)]{Lea2010}
{Luna, G.J.M., Sokoloski, J., Mukai, K., \& Nelson, T.}, 2010,
\textit{ATel}, 3053

\bibitem[Nelson \etal\ (2011)]{Nea2011}
{Nelson, T., Mukai, K., Orio, M., Luna, G.J.M., \& Sokoloski, J.L.}
2011, \textit{ApJ}, 737, A7

\bibitem[Pagnotta \etal\ (2009)]{Pea2009}
{Pagnotta, A., Schaefer, B.E., Xiao, L., Collazzi, A.C., \& Kroll, P.}
2009, \textit{AJ}, 138, 1230

\bibitem[Patterson \& Raymond (1985)]{PR1985}
{Patterson, J., \& Raymond, J.C.} 1985, \textit{ApJ}, 292, 535

\bibitem[Schaefer (2010)]{S2010}
{Schaefer, B.E.} 2010, \textit{ApJSupp}, 187, 275

\bibitem[Selvelli \etal\ (1992)]{Sea1992}
{Selvelli, P.L., Casatella, A., \& Gilmozzi, R.} 1992, \textit{ApJ}, 393, 289

\bibitem[Takei \etal\ (2011)]{Tea2011}
{Takei, D., Ness, J.U., Tsujimoto, M., Kitamoto, S., Drake, J.J.,
Osborne, J.P., Takahashi, H., \& Kinugasa, K.} 2011, \textit{PASJ}, in press

\end{thebibliography}
\end{document}